\documentclass[aip, jcp, reprint]{revtex4-1}

\usepackage{enumitem}
\usepackage{graphicx}          
\usepackage{dcolumn}           
\usepackage{bm}                
\usepackage[mathlines]{lineno} 
\usepackage[utf8]{inputenc}
\usepackage{mathrsfs}
\usepackage{amsmath}    
\usepackage{amssymb}    
\usepackage{amsthm}
\usepackage{graphicx}   
\usepackage{verbatim}   
\usepackage{color}      
\usepackage{subfigure}  
\usepackage{hyperref}   
\usepackage[capitalise]{cleveref}   
\usepackage{appendix}
\DeclareMathOperator*{\argmax}{arg\,max}

\newcommand{\vech}{\operatorname{vech}}

\DeclareMathOperator{\Tr}{Tr}

\DeclareMathOperator{\diag}{diag}

\begin{document}

\title{Efficient maximum likelihood parameterization of continuous-time Markov processes}
\author{Robert T. McGibbon}
\affiliation{Department of Chemistry, Stanford University, Stanford CA 94305, USA}
\author{Vijay S. Pande}
\affiliation{Department of Chemistry, Stanford University, Stanford CA 94305, USA}
\email{pande@stanford.edu}
\date{\today}

\begin{abstract}
Continuous-time Markov processes over finite state-spaces are widely used to model dynamical processes in many fields of natural and social science. Here, we introduce an maximum likelihood estimator for constructing such models from data observed at a finite time interval. This estimator is dramatically more efficient than prior approaches, enables the calculation of deterministic confidence intervals in all model parameters, and can easily enforce important physical constraints on the models such as detailed balance. We demonstrate and discuss the advantages of these models over existing discrete-time Markov models for the analysis of molecular dynamics simulations.
\end{abstract}
\maketitle

\section{Introduction}
Estimating the parameters of a continuous-time Markov jump process model based on discrete-time observations of the state of a dynamical system is a problem which arises in many fields of science, including physics, biology, sociology, meteorology, and finance.\cite{singer1976representation, madsen1985markov, turner1989markov, minin2008counting} Diverse applications include the progression of credit risk spreads,\cite{jarrow1997markov} social mobility,\cite{spilerman1972extensions} and the evolution of DNA sequences in a phylogenetic tree.\cite{kimura1980simple} In chemical physics, these models, also called master equations, describe first-order chemical kinetics, and are the principle workhorse for modeling chemical reactions.\cite{anderson2011continuous}

For complex physical systems, the derivation of kinetic models from first principles is often intractable. In these circumstances, the parameterization of models from data is often a superior approach. As an example, consider the
dynamical behavior of solvated biomolecules, such as proteins and nucleic acids. Despite the microscopic complexity of their equations of motion, relatively simple multi-state kinetics often arise, as exemplified by the ubiquity of two- and few-state Markov process models for protein folding.
\cite{ikai1971kinetic, zwanzig1997two, Sanchez2003367, chan1998protein, pirchi2011single, beauchamp2012simple}

Due in part to the unavailability of computationally efficient and numerically robust estimators for continuous-time Markov models, in the field computational biophysics, discrete-time Markov models have been widely used to fit and interpret the output of molecular dynamics (MD) simulations. Also called Markov state models (MSMs), these methods describe the molecular kinetics observed in an MD simulation as a jump process with a discrete-time interval generally on the order of $\sim$ 10 -- 100 ns.\cite{Beauchamp2011Msmbuilder2, senne2012emma} These models provide convenient estimators for key quantities of interest for molecular systems, such as the free energies of various metastable conformational states, the timescales of their interconversion, and the dominant transition pathways.\cite{banerjee2014transition, shukla2015markov, noe2008transition, chodera2014markov}

In this work, we introduce an efficient maximum likelihood estimator for
continuous-time Markov models on a finite state space from discrete-time data.
The source of data used here is identical to that employed in fitting discrete-time Markov chain models -- namely, the number of observed transitions between each pair of states within a specified time interval. We demonstrate the properties of these models on simple systems, and apply them to the analysis of the folding of the FiP35 WW protein domain.

\section{Background}
\label{s:background}
Consider a time-homogenous continuous-time Markov process $\{X(t): t \geq 0\}$ over a finite state space, $\mathscr{S}=\{1, \ldots, n\}$. The process is determined completely by an $n \times n$ matrix $\mathbf{K}$, variously called its rate matrix, infinitesimal generator,\cite{Metzner2007353} substitution matrix,\cite{holmes2002expectation} or intensity matrix.\cite{Bladt2005Statistical}

For an interval $\tau>0$, begin with the $n \times n$ matrix, $\mathbf{T}(\tau)$, of probabilities that the process jumps from one state, $i$, to another state, $j$,
\begin{align}
\mathbf{T}(\tau)_{ij} = P(X(t+\tau) = j \;|\; X(t) = i),
\end{align}
which, by time-homogeneity is assumed to be independent of $t$. The process's rate matrix, $\mathbf{K}$, is defined as
\begin{align}
\mathbf{K} \equiv \lim_{\tau \rightarrow 0+} \frac{\mathbf{T}(\tau) - \mathbf{I}_{n}}{\tau}.
\label{eq:def_K}
\end{align}

Given $\mathbf{K}$ and any time interval, $\tau$, the transition probability matrix, $\mathbf{T}(\tau)$, can be expressed a matrix exponential
\begin{align}
\mathbf{T}(\tau) = \exp(\mathbf{K}\tau) \equiv \sum_{i=0}^\infty \frac{\tau^i \mathbf{K}^i}{i!}.
\end{align}

A particular rate matrix $\mathbf{K}$ corresponds to a valid continuous-time Markov process if and only if its off-diagonal elements are nonnegative and its row sums equal zero. These constraints are necessary to ensure that the probabilities propagated by the dynamics remain positive and sum to one. We denote by $\mathscr{K}$ this set of admissible rate matrices,
\begin{align}
\label{eq:4}
 \mathscr{K} = \Bigg\{\mathbf{K} = \{k_{ij}\} \in \mathbb{R}^{n \times n} : k_{ij} &\geq 0 \text{ for all } i \neq j, \\\nonumber k_{ii} &= -\sum_{j\neq i} k_{ij} \Bigg\}.
\end{align}

Furthermore, we denote by $\mathscr{T}$ the set of all embeddable transition probability matrices, that is, those which could originate as the transition probability matrix, $\mathbf{T}(\tau)$, induced by some continuous-time Markov process,
\begin{align}
\mathscr{T} = \left\{\mathbf{T} \in \mathbb{R}^{n \times n} : \exists\; \mathbf{K} \in \mathscr{K} s.t. \mathbf{T}=\exp(\mathbf{K}) \right\}.
\end{align}

It is well-known that set $\mathscr{T}$ is a strict subset of the set of all stochastic matrices; not all stochastic matrices are embeddable.\cite{kingman1962imbedding, davies2010embeddable} A complete description of the topological structure of $\mathscr{T}$ as well as the necessary and sufficient conditions for a stochastic matrix to be embeddable are
open problems in the theory of Markov processes.

Although \cref{eq:def_K} serves as the definition of the rate matrix of a continuous-time Markov process, it is generally not directly suitable as a method for parameterizing Markov models, particularly for applications in chemical kinetics. The attempt to numerically approximate the limit in
\cref{eq:def_K} from empirically measured transition probabilities would be valid if the generating process were exactly Markovian. However, in chemical kinetics, a Markov process model -- the chemical master equation -- is an approximation valid only for timescales longer than the molecular relaxation time.\cite{adams1981dynamical,voter1985dynamical} A suitable Markov model which is predictive over long timescales must capture both the instantaneous kinetics as well as, to use the vocabulary of Mori-Zwanzig formalism, the effective contribution of the integrated memory kernel.\cite{mori1965transport, zwanzig1961memory}


Our goal is to address this parameterization problem. The primary contribution of this work is an efficient algorithm for estimating $\mathbf{K}$ from observed discrete-time observations. We adopt a direct maximum likelihood approach, with $O(n^3)$ work per iteration. Many constraints on the solution, such as detailed balance or specific sparsity patterns on $\mathbf{K}$ can be introduced in a straightforward manner without additional cost.

Prior work on this subject is numerous. Crommelin and Vanden-Eijnden proposed a method for estimating $\mathbf{K}$ in which a discrete-time transition probability matrix is first fit to the observed data, followed by the determination of the rate matrix, $\mathbf{K}$ such that $\exp(\mathbf{K}\tau)$ is nearest to the target empirical transition probability matrix.\cite{crommelin2006fitting, Crommelin2009Databased} The nature of this calculation depends on the norm used to define the concept of ``nearest'': under a Frobenius norm, this problem has a closed form solution, while the norm of Crommelin and Vanden-Eijnden leads to a quadratic program. A similar approach was advocated by Israel et al.\cite{israel2001finding}

Kalbfleisch and Lawless proposed a maximum likelihood estimator for $\mathbf{K}$.\cite{Kalbfleisch1985analysis} Without constraints on the rate matrix, their proposed optimization involves the construction and inversion of an $n^2 \times n^2$ Hessian matrix at each iteration of the optimization, rendering it prohibitively costly ($O(n^6)$ scaling per iteration) for moderate to large state spaces.

A series of expectation maximization (EM) algorithms are described by Asmussen, Nerman and Olsson, Holmes and Rubin, Bladt and S{\o}rensen, and Hobolth and Jensen \cite{asmussen1996fitting, holmes2002expectation, Bladt2005Statistical, hobolth2005statistical} These algorithms treat the state of the system between observation intervals as an unobserved latent variable, which when interpolated via EM leads to more efficient estimators. A review of these algorithms is presented by Metzner et al.\cite{Metzner2007353} At best, each iteration of the proposed methods scales as $O(n^5)$.

\section{Maximum likelihood estimation}
\subsection{Log-likelihood and gradient}

We take our source of data to be one or more observed discrete-time trajectories from a Markov process, $x = \{x_{0}, x_{\tau}, \ldots, x_{N\tau} \}$, in a finite state space, observed at a regular time interval.

The likelihood of the data given the model and the initial state is given in terms of the transition probability matrix as the product of the transition probabilities assigned to each of the observed jumps in the trajectory.

\begin{align}
\label{eq:8}
P(x | \mathbf{K}, x_0) = \prod_{k=0}^{N-1} \mathbf{T}(\tau)_{x_{k\tau},\, x_{(k+1)\tau}}.
\end{align}
When more than one independent trajectory is observed, the data likelihood is a product over trajectory with individual terms given by \cref{eq:8}.

Because many transitions are potentially observed multiple times, \cref{eq:8} generally contains many repeated terms. Define the observed transition count matrix $\mathbf{C}(\tau) \in \mathbb{R}^{n \times n}$,
\begin{align}
\mathbf{C}(\tau)_{ij} = \sum_{k=0}^{N-1} \mathbf{1}(x_{k\tau} = i) \cdot \mathbf{1}(x_{(k+1)\tau} = j).
\end{align}
Collecting repeated terms, the likelihood can be rewritten more compactly as
\begin{align}
P(x | \mathbf{K}, x_0) = \prod_{i,j} \mathbf{T}(\tau)_{ij}^{\mathbf{C}(\tau)_{ij}}.
\end{align}

Suppose that the rate matrix, $\mathbf{K}$ is parameterized by a vector, $\theta \in \mathbb{R}^b$ of independent variables, $\mathbf{K}=\mathbf{K}(\theta)$.  In the most general case, every element of the rate matrix may individually be taken as an independent variable, with $b=n^2 - n$. As discussed in \cref{ss:denseparam}, other parameterizations may be used to enforce certain properties on $\mathbf{K}$. The logarithm of data likelihood is
\begin{align}
\mathcal{L}(\theta; \tau) &\equiv \ln P(x | \mathbf{K}(\theta), x_0), \\
&= \sum_{i,j} \mathbf{C}_{ij}(\tau) \ln \mathbf{T}(\tau)_{ij}, \label{eq:loglt} \\
&= \sum_{i,j} \Big(\mathbf{C}(\tau) \circ \ln \exp\big(\tau\, \mathbf{K}(\theta)\big)\Big)_{ij}, \label{eq:logl}
\end{align}
where $\ln(\mathbf{X})$ is the element-wise natural logarithm, $\exp(\mathbf{X})$ matrix exponential, and $\mathbf{X} \circ \mathbf{Y}$ is the Hadamard (element-wise) matrix product. Note that the element-wise logarithm and matrix exponential are not inverses of one another.

The most straightforward parameter estimator -- the maximum likelihood estimator (MLE) -- selects parameters which maximize the likelihood of the data,
\begin{align}
\label{eq:mle}
\theta^{MLE} = \argmax_{\theta}\, \mathcal{L}(\theta; \tau).
\end{align}

To maximize \cref{eq:mle}, we focus our attention on quasi-Newton optimizers that utilize the first derivatives of $\mathcal{L}(\theta; \tau)$ with respect to $\theta$. This requires an efficient algorithm for computing $\nabla_\theta \mathcal{L}(\theta; \tau)$. We achieve this by starting from the eigendecomposition of $\mathbf{K}$,
\begin{align}
\mathbf{K} = \mathbf{V}\diag(\lambda)\mathbf{U}^T,
\end{align}
where the columns of $\mathbf{U}$ and $\mathbf{V}$ contain the left and right eigenvectors of $\mathbf{K}$ respectively, jointly normalized such that $\mathbf{V}^{-1} = \mathbf{U}^T$, and $\lambda$ are the corresponding eigenvalues. Assuming that $\mathbf{K}$ has no repeated eigenvalues, the directional derivatives of induced transition probability matrix, $\partial \mathbf{T}(\tau)_{ij} / \partial\theta_u$ are given by \cite{Kalbfleisch1985analysis, Jennrich1976fitting}
\begin{align}
\partial\mathbf{T}(\tau)_{ij}/\partial\theta_u = \mathbf{V}\left(
(\mathbf{U}^T(\partial\mathbf{K}/\partial\theta_u)\mathbf{V})
\circ \mathbf{X}(\lambda, t)\right)\mathbf{U}^T,
\end{align}
where $\mathbf{X}(\lambda, t)$ is an $n \times n$ matrix with entries
\begin{align}
\left[\mathbf{X}(\lambda, t)\right]_{ij} = \begin{cases}
\tau\; \exp(\tau \lambda_i), & i=j, \\
\frac{\exp(\tau \lambda_i) - \exp(\tau \lambda_j)}{\lambda_i - \lambda_j}, & i \neq j.
\end{cases}\label{eq:X}
\end{align}

The elements of the gradient of the log-likelihood can then be constructed as
\begin{align}
\label{eq:naivegradient}
\frac{\partial \mathcal{L}(\theta;\tau)}{\partial \theta_u} = \sum_{ij}\left(
\mathbf{D} \circ \mathbf{V} \Big(
\left(\mathbf{U}^{T} (\partial \mathbf{K}/\partial \theta_u) \mathbf{V}\right) \circ \mathbf{X}(\lambda, t)
\Big)
\mathbf{U}^{T}\right)_{ij},
\end{align}
where $\mathbf{D}_{ij} = \mathbf{C}(\tau)_{ij} / \mathbf{T}_{ij}$.

A direct implementation of \cref{eq:naivegradient} requires at least 4 $n \times n$ matrix multiplies for each element of $\theta$, indexed by $u$. If the parameter vector, $\theta$, contains $O(n^2)$ parameters, then computing the full gradient will require $O(n^5)$ floating point operations (FLOPs). However, two properties of the Hadamard product and matrix trace can be exploited to dramatically reduce the computational complexity of constructing the gradient vector to $O(n^3)$ FLOPs.

\begin{align}
\sum_{ij}(\mathbf{A}\circ \mathbf{B})_{ij} &= \Tr\left(\mathbf{A}\mathbf{B}^T\right), \\
\Tr\left(\mathbf{A}^T (\mathbf{B} \circ \mathbf{C})\right) &= \Tr\left(\mathbf{B}^T (\mathbf{A} \circ \mathbf{C})\right).
\end{align}

Using these identities, the gradient of the log-likelihood can be rewritten as
\begin{align}
\label{eq:efficientgradient}
\frac{\partial \mathcal{L}(\theta;\tau)}{\partial \theta_u} = \sum_{ij} \Bigg( \partial \mathbf{K}/d\theta_u \circ \underbrace{\left(\mathbf{U} \Big( (\mathbf{V}^{T} \mathbf{D} \mathbf{U}) \circ \mathbf{X}(\lambda, t) \Big) \mathbf{V}^T\right)}_{\mathbf{Z}} \Bigg)_{ij}.
\end{align}

Note that because $\mathbf{Z}$ is independent of $u$, it can be constructed once at the beginning of a gradient calculation at a cost of $O(n^3)$ FLOPs, and reused for each index, $u$. The remainder of the work involves constructing the derivative matrix $\partial\mathbf{K}/\partial\theta_u$, which is generally quite sparse, and a single inexpensive sum of a Hadamard product. Overall, this rearrangement reduces the complexity of constructing the full gradient vector from $O(n^5)$ to $O(n^3)$ FLOPs.

\subsection{Reversible parameterization}
\label{ss:denseparam}
In the application of these models to domain-specific problems, additional constraints on the Markov process may be known, and enforcing these constraints during parameterization can enhance the interpretability of solutions as well as provide a form of regularization.

For many molecular system, it is known that the underlying dynamics are reversible, and this property can be enforced in Markov models as well. A Markov process is reversible when the rate matrix, $\mathbf{K}$, satisfies the detailed balance condition with respect to a stationary distribution, $\pi$, towards which the process relaxes over time.
\begin{align}
\pi \mathbf{K} &= 0, \\
\label{eq:rev}
\pi_i k_{ij} &= \pi_j k_{ji} \hspace{2em} \forall\; i \neq j.
\end{align}

This constraint can be enforced on solutions through the design of the parameterization function, $\mathbf{K}(\theta)$. If $\mathbf{K}$ is reversible, \cref{eq:rev} implies that a real symmetric $n \times n$ matrix, $\mathbf{S}$, can be formed, which we refer to as the symmetric rate matrix, such that
\begin{align}
\mathbf{S} = \mathbf{S}^T = \diag(\sqrt{\pi}) \,\mathbf{K}\diag(\sqrt{\pi})^{-1}. \label{eq:symrate}
\end{align}

Because of this symmetry and the constraint on the row sums of $\mathbf{K}$, only the upper triangular (exclusive of the main diagonal) elements of $\mathbf{S}$, and the stationary vector, $\pi$, need be directly encoded by the parameter vector, $\theta$, to fully specify $\mathbf{K}$. Furthermore, since the elements of $\pi$ are constrained to be positive, working with the element-wise logarithm of $\pi$ can enhance numerical stability. For the elements of $\mathbf{S}$, which are only constrained to be nonnegative, the same logarithm transformation is inapplicable, as it is incompatible with sparse solutions that set one or more rate constants equal to zero. For these reasons, we use a parameter vector, of length $b=\binom{n+1}{2}$, with $\theta=(\theta^{(S)}, \theta^{(\pi)})$. The first $\binom{n}{2}$ elements, notated $\theta^{(S)}$, encode the off-diagonal elements of $\mathbf{S}$. The remaining $n$ elements are notated $\theta^{(\pi)}$, and are used to construct the stationary distribution, $\pi$. From $\mathbf{S}$ and $\pi$, the off-diagonal and diagonal elements of $\mathbf{K}$ are then constructed from \cref{eq:symrate}. In explicit notation, the construction is
\begin{align}
\vech(\mathbf{S})_i &= \theta^{(S)}_i \hspace{1em} i \in \{1, \ldots, n(n-1)/2\}, \\
\pi_i &= \frac{\exp(\theta^{(\pi)}_i)}{\sum_{j=1}^n \exp(\theta^{(\pi)}_j)}   \hspace{1em} i \in \{1, \ldots, n\}, \label{eq:pii} \\
\mathbf{K}_{ij} &= \begin{cases}
[D(\sqrt{\pi})^{-1} \mathbf{S} D(\sqrt{\pi})]_{ij}, & i \neq j\\
-\sum_{j\neq i}\mathbf{K}_{ij}, & i = j,
\end{cases} \label{eq:kij}
\end{align}
where $\vech(\mathbf{A})$ is the row-major vectorization of the elements of a symmetric $n \times n$ matrix above the main diagonal,
\begin{align}
\vech(\mathbf{A}) = [a_{1,2}, \ldots, a_{1,n}, a_{2,3},\ldots,a_{2,n},\ldots, a_{n-1, n} ]^T.
\end{align}

The necessary gradients of \cref{eq:kij}, $\partial\mathbf{K}_{ij}/d\theta_u$ are sparse. For fixed $1 \leq u \leq \binom{n}{2}$, the $n \times n$ matrix $\partial\mathbf{K}_{ij}/d\theta_u$ over all $i$, $j$ contains only four nonzero entries, whereas for $\binom{n}{2} < u \leq \binom{n+1}{2}$, the same matrix contains $3n-2$ nonzero entries. The sum of its Hadamard product with $\mathbf{Z}$ in \cref{eq:efficientgradient} can thus be computed in $O(1)$ or $O(n)$ time. For the remainder of this work, we focus exclusively on this reversible parameterization for $\mathbf{K}(\theta)$.


\subsection{Optimization}

Equipped with the log-likelihood and an efficient algorithm for the gradient, we now consider the construction of maximum likelihood estimates, \cref{eq:mle}. Among the first-order quasi-Newton methods tested, we find Limited-memory Broyden-Fletcher-Goldfarb-Shanno optimizer with bound constraints (L-BFGS-B) to be the most successful and robust.\cite{byrdlimited1995, zhu1997Algorithm}

To begin the optimization, we choose the initial guess for $\theta$ according to the following procedure. First, we fit the maximum likelihood reversible transition probability matrix computed using Algorithm 1 of \citet{prinz2011markov}. Next, we compute its principle matrix logarithm, $\widetilde{\mathbf{K}}$, using an inverse scaling and squaring algorithm, and scaling by $\tau$.\cite{AlMohy2012Improved} Generally, the MLE reversible transition matrix is not embeddable, and thus the principle logarithm is complex or has negative off-diagonal entries, and does not correspond to any valid continuous-time Markov process. We take the initial guess from $\theta^{(\pi)}$ directly from the stationary eigenvector of the MLE transition matrix, and $\theta^{(S)}$ from the nearest (by Frobenius norm) valid rate matrix to $\widetilde{\mathbf{K}}$, given by $\max(\mathrm{Re}(\widetilde{\mathbf{K}}),0)$. \cite{davies2010embeddable}

The optimization problem is non-convex in the general case and may have multiple local minima. Varying the optimizer's initialization procedure can thus mitigate the risk of convergence to a low quality local minimum. One alternative initialization $\mathbf{K}$ is the pseudo-generator, $\mathbf{K}_p = (\mathbf{T}(\tau)-I_n) / \tau$, which arises from a first-order Taylor approximation to the matrix exponential. After the optimization has terminated, a useful check is to compare the maximum likelihood transition matrix $\mathbf{T}(\tau)$ estimated during initialization with the exponential of the recovered rate matrix, $\exp(\tau\mathbf{K}^{MLE})$. Large differences between the two matricies, or their eigenspectra / relaxation timescales, may be symptomatic of non-embedability of the data or a convergence failure of the optimizer. If the data are available at a lag time shorter than $\tau$, convergence failures can often also be circumvented by using a converged rate matrix obtained from a model at a shorter lag time as an initial guess for a model at a longer lag time.

\subsection{Implementation notes}
Because $\mathbf{S}$ is symmetric, it can be diagonalized efficiently at cost of $O(4n^3/3)$ FLOPs. The eigenvectors can then be rotated by $D(\sqrt{\pi})$ to give the eigenvectors of $\mathbf{K}$. Compared to diagonalizing the non-symmetric matrix $\mathbf{K}$ directly, this can yield a speedup of 2-10$\times$ in the critical diagonalization step required to compute the gradient vector.

For each pair of states with an observed transition count, $(i, j) \text{ such that } \mathbf{C}(\tau)_{ij}>0$, the gradient expressions \cref{eq:naivegradient}, and \cref{eq:efficientgradient} are only defined when $\mathbf{T}_{ij} > 0$. A sufficient condition to ensure this property is that $\mathbf{K}$ be irreducible,\cite{bakry2013analysis} but this cannot be straightforwardly ensured throughout every iteration of the L-BFGS-B optimization without heavy-handed measures such as complete positivity of $\mathbf{K}$. In practice, we find that replacing any zeros values in $\mathbf{T}$ with a small constant, such as $1 \times 10^{-20}$, when computing the matrix $\mathbf{D}$ in \cref{eq:efficientgradient} is sufficient to avoid this instability.

Furthermore, note that calculation of $\mathbf{X}(\lambda, t)$ by direct implementation of \cref{eq:X} can suffer from a substantial loss of accuracy for close-lying eigenvalues. The matrix can instead be computed in a more precise manner using the $\texttt{exprel}(x) \equiv (e^x - 1) / x$ or $\texttt{exmp1} \equiv e^x - 1$ routines, which are designed to be accurate for small $x$ and are available in numerical libraries such as SLATEC, GSL, and the upcoming release of SciPy.\cite{vandevender1982slatec, gough2009gnu, scipy}

\section{Quantifying Uncertainty}
\label{sect:uncertainty}
Since all data sets are finite, statistical uncertainty in any estimate of a probabilistic model is unavoidable. Therefore, key quantities of interest beyond the maximum likelihood rate matrix itself, $\mathbf{K}^{MLE} = \mathbf{K}(\theta^{MLE})$, are estimates of the sampling uncertainty in $\mathbf{K}^{MLE}$, and estimates of the sampling uncertainty in quantities derived from $\mathbf{K}^{MLE}$, such as its stationary eigenvector, $\pi$, its eigenvalues, $\lambda_i$, and relaxation timescales.


In the large sample size limit, the central limit theorem guarantees that the distribution of $\theta^{MLE}$ converges to a multivariate normal distribution with a covariance matrix which can be estimated by the inverse of the Hessian of the log-likelihood function evaluated at $\theta^{MLE}$, assuming that the MLE does not lie on a constraint boundary.\cite{rao2009linear} This can be thought of as a second order Taylor expansion for the log-likelihood surface at the MLE; the log-likelihood is approximated as a paraboloid with negative curvature whose peak is at the MLE and whose width is determined by the Hessian matrix at the peak. The exponential of the log-likelihood, the likelihood surface, is then Gaussian, and the multivariate delta theorem can be used to derive expressions for the asymptotic variance in scalar functions of $\theta^{MLE}$.\cite{rao2009linear} Computationally, the critical component is the computation of the Hessian matrix,
\begin{align}
&\mathbf{H}_{uv}(\theta; \tau) = \frac{\partial^2 \mathcal{L}(\theta;\tau)}{\partial\theta_u \partial\theta_v}, \\
&= \sum_i^n\sum_j^n \mathbf{C}_{ij} \left( \frac{\partial^2 \mathbf{T}_{ij} / \partial \theta_u \partial \theta_v}{\mathbf{T}_{ij}} - \frac{(\partial \mathbf{T}_{ij} / \partial\theta_u)(\partial \mathbf{T}_{ij} / \partial\theta_v)}{\mathbf{T}_{ij}^2} \right). \label{eq:hessian}
\end{align}
and its inverse.

\subsection{Approximate Analytic Hessian}
Direct calculation of the Hessian requires both the evaluation of the first derivatives of $\mathbf{T}$ as well as the more costly second derivatives. A more efficient alternative, as pointed out by Kalbfleisch and Lawless, is to approximate the second derivatives by estimates of their expectations.\cite{Kalbfleisch1985analysis}

Let $C_i = \sum_{j}\mathbf{C}_{ij}$. Taking the expected value of $\mathbf{C}_{ij}$ conditional on $C_i$, we approximate $\mathbf{C}_{ij} \approx \mathbf{T}_{ij} C_i$. This makes it possible to factor $\mathbf{C}_{ij}$ out of the summation over $j$ in \cref{eq:hessian}, and exploit the property that $\sum_{j}^n \partial^2 \mathbf{T}_{ij} / \partial \theta_u \partial \theta_v = 0$, simplifying \cref{eq:hessian} to
\begin{align}
&\mathbf{H}_{uv}(\theta; \tau) \approx - \sum_{ij} \frac{C_i}{\mathbf{T}_{ij}} \frac{\partial \mathbf{T}_{ij}}{\partial\theta_u}
\frac{\partial \mathbf{T}_{ij}}{\partial\theta_v}. \label{eq:approxhessian}
\end{align}

Equipped with the approximator \cref{eq:approxhessian}, the asymptotic variance-covariance matrix of $\theta$ is calculated as the matrix inverse of the Hessian, $\mathbf{\Sigma} = \mathbf{H}^{-1}$, and the asymptotic variance in each derived quantity $g(\theta)$ is estimated using the multivariate delta method.\cite{rao2009linear}
\begin{align}
\text{Var}(g(\theta)) \approx \nabla g(\theta^{MLE})^T \,\mathbf{\Sigma}\, \nabla g(\theta^{MLE})
\end{align}

For example, the asymptotic variance in the stationary distribution can be calculated as
\begin{align}
\text{Var}(\pi_k) \approx \sum_{i,j}^n \frac{\partial \pi_k}{\partial \theta^{(\pi)}_i} \mathbf{\Sigma}_{ij}^{(\pi)} \frac{\partial \pi_k}{\partial \theta^{(\pi)}_j},
\end{align}
where $\mathbf{\Sigma}^{(\pi)}$ represent the lower $n \times n$ block of the asymptotic variance covariance matrix and
\begin{align}
\frac{\partial \pi_i}{\partial \theta^{(\pi)}_j} &= \begin{cases}
\pi_i -\pi_i^2, & i=j, \\
- \pi_i \pi_j, & i \neq j,
\end{cases}
\end{align}

Other key quantities of interest for biophysical applications include the exponential relaxation timescales of the Markov model
\begin{align}
\label{eq:34}
\tau_i = -(\lambda_i)^{-1} \hspace{1em} i \in \{2, \ldots, n\}.
\end{align}
The asymptotic variance in the relaxation timescales, $\tau_i$, is
\begin{align}
\text{Var}(\tau_i) &\approx \sum_{uv} \frac{\partial \tau_i}{\partial \theta_u} \Sigma_{uv} \frac{\partial \tau_i}{\partial \theta_v},
\end{align}
where $\frac{\partial \tau_i}{\partial \theta_u}$ follows from standard expressions for derivatives of eigensystems,\cite{murthy1988derivatives}
\begin{align}
\frac{\partial \tau_i}{\partial \theta_u} = \frac{1}{\lambda_i^2} \left[ \mathbf{U}^T \frac{\partial \mathbf{K}(\theta)}{\partial \theta_u} \mathbf{V} \right]_{ii}.
\end{align}

The sampling uncertainty in other derived properties which depend continuously on $\theta$ can be calculated similarly.

When the MLE solution lies at the boundary of the feasible region, with one or more elements of $\theta^{(S)}$ equal to zero, we adopt an active set approach to approximate $\mathbf{\Sigma}$. We refer to the elements of $\theta^{(S)}$ which do not lie on a constraint boundary as free parameters. The Hessian block for the free parameters is constructed and inverted, and the variance and covariance of the constrained elements as well as their covariance with the free parameters is taken to be zero.

\section{Numerical Experiments}
We performed numerical experiments on three datasets, which demonstrate different aspects of our estimator for continuous-time Markov processes. Where appropriate, we compare these models to reversible discrete-time Markov models which directly estimate $\mathbf{T}(\tau)$, parameterized via Algorithm 1 of \citet{prinz2011markov}

\begin{figure}
\includegraphics[width=2in]{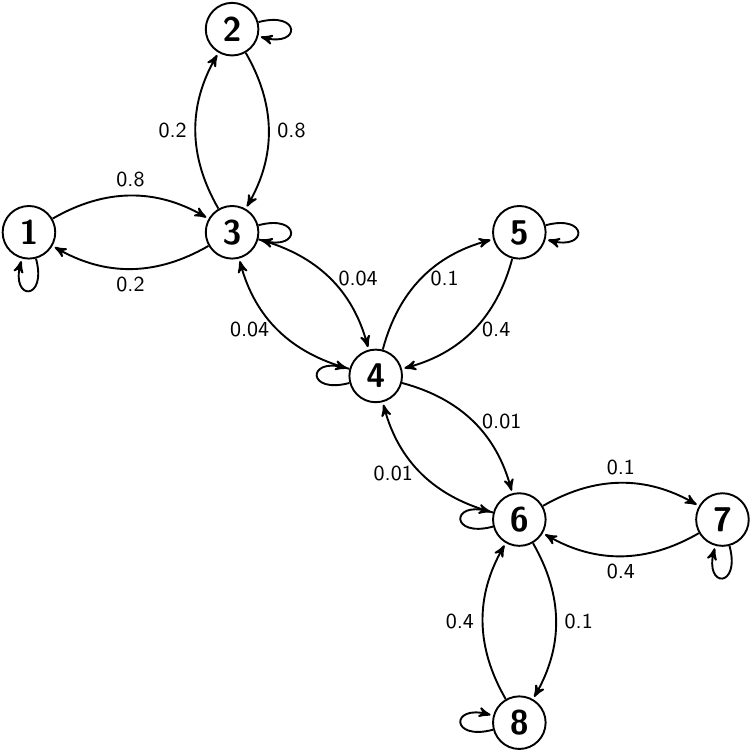}
\caption{\label{fig:network} A simple eight state Markov process. Connected states are labeled with the pairwise rate constants, $\mathbf{K}_{ij}$. Self transition rates (not shown), $\mathbf{K}_{ii}$, are equal to the negative sum of each states outgoing transition rates, in accordance with \cref{eq:4}.}
\end{figure}

\begin{figure}
\includegraphics[width=3in]{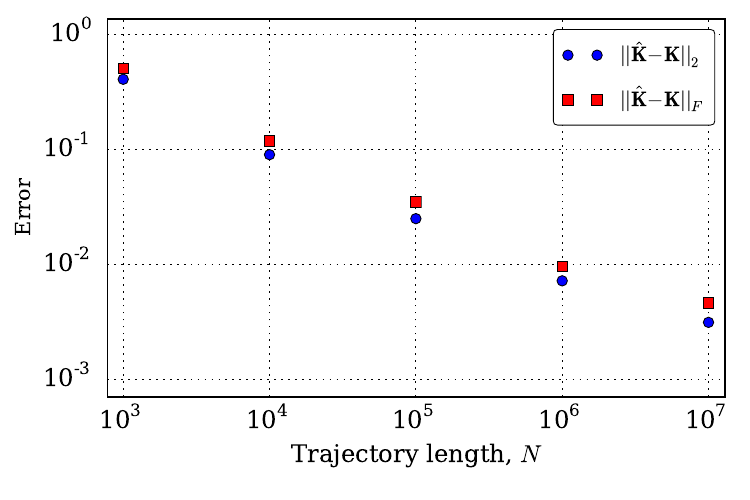}
\caption{\label{fig:convergence}Convergence of the estimated rate matrix, $\mathbf{\hat{K}}$, to the true generating rate matrix in \cref{fig:network} for discrete-time trajectories of increasing length simulated from the process in \cref{fig:network} with a time step of 1. Using either a 2 norm (blue) or Frobenius norm (red), we see roughly power law convergence over the range of trajectory lengths studied.}
\end{figure}

\begin{figure}
\includegraphics[width=3in]{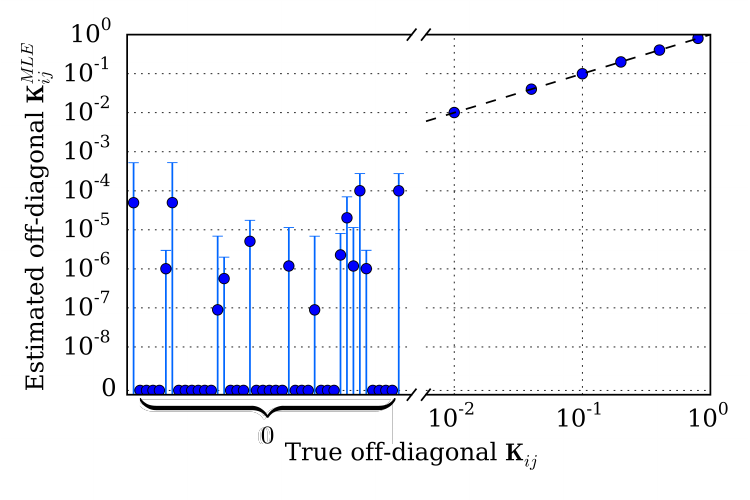}
\caption{\label{fig:network_comp}Comparison of the estimated and true off-diagonal rate matrix elements for a trajectory of length $N=10^{7}$ simulated from the process in \cref{fig:network} with a time step of 1. The true non-zero elements of $\mathbf{K}$ are well-estimated, as shown in the right portion of the plot; here, error bars are small enough to be fully obscured by point markers.
On the other hand, the estimator spuriously estimates non-zero rates between many of the states which are not connected in the underlying process. However, the 95\% confidence intervals for these spurious rates each overlap with zero.}
\end{figure}

\subsection{Recovering a Known Rate Matrix}
\label{exp:1}
First, we constructed a simple synthetic eight state Markov process with known rates. The network is shown in \cref{fig:network}. The largest non-zero eigenvalue of $\mathbf{K}$ is $\lambda_2 \approx -9.40 \times 10^{-3}$, which corresponds to a slowest exponential relaxation timescale, $\tau_2 \approx 106.4$ (arbitrary time units).

From this model, we simulated discrete-time data with a collection interval of 1 time unit by calculating the matrix exponential of $\mathbf{K}$ and propagating the discrete-time Markov chain. In \cref{fig:convergence}, we show
the convergence of the models estimated from this simulation data to the true model, as the length of the simulated trajectories grows. As expected, the fit parameters get more accurate as the size of the data set grows. We observe approximately power law convergence as measured by the 2-norm and Frobenius norm over the range of trajectory lengths studied.

The true rate matrix for this continuous time Markov process is sparse -- only 7 of the 28 possible pairs of distinct states are directly connected in \cref{fig:convergence}. Can this graph structure be recovered by our estimator?
This task is challenging because of the nature of the discrete-time data. The observation that the system transitioned from state $i$ (at time $t$) to state $j$ (at time $t+1$) does \emph{not} imply that $\mathbf{K}_{ij}$ is non-zero. Instead, the observed $i\rightarrow j$ transition may have been mediated by one or more other states -- the process may have jumped from $i$ to $k$, and then again from $k$ to $j$, all within the observation interval.

When the rate matrix, $\mathbf{K}$ is irreducible, the corresponding transition probability matrix $\mathbf{T}(\tau)$ is strictly positive for every positive lag time, $\tau$.\cite{bakry2013analysis} This implies that in the limit that the trajectory length, $N$, approaches infinity, at least 1 transition count will almost surely be observed between any pair of states, regardless of the sparsity of $\mathbf{K}$.

In \cref{fig:network_comp}, we attempt to resolve the underlying graph structure using the model estimated with a trajectory of length $N=10^7$. The plot compares the estimated rate matrix elements with the true values. We find that all of the true connections are well-estimated, and that many of the zero rates are also correctly identified. However, the maximum likelihood estimator also identifies very low, but non-zero rates between many of the states which are in fact disconnected.

We computed 95\% ($1.96\sigma$) confidence intervals for each of the estimated rate matrix elements, $\mathbf{K}^{MLE}_{ij}$. For each of the spuriously non-zero elements, these confidence intervals overlapped with zero. None of the confidence intervals for the properly non-zero rates overlapped with zero. These uncertainty estimates can therefore be used, in combination with the MLE,
to identify the underlying graph structure.

This example demonstrates that some degree of sparsity-inducing regularization or variable selection may be required to robustly identify the underlying graph structure in Markov process. 

\subsection{Accuracy of Uncertainty Estimates}
\label{ss:numerical}
\begin{figure}
\includegraphics[width=3in]{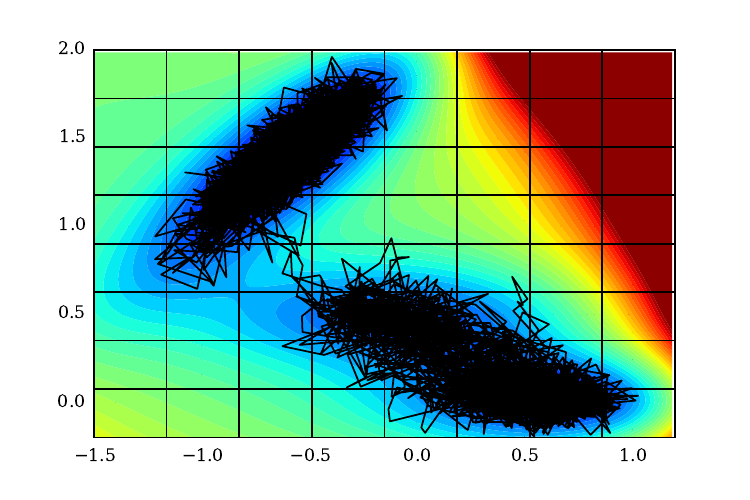}
\caption{\label{fig:muller}Brownian dynamics on the 2-dimensional M\"uller potential was discretized by projecting the simulated trajectories onto an $8\times8$ grid. A typical trajectory is shown in black. The resulting discrete-state process can be approximated as a continuous-time Markov process.}
\end{figure}

\begin{figure}
\includegraphics[width=3in]{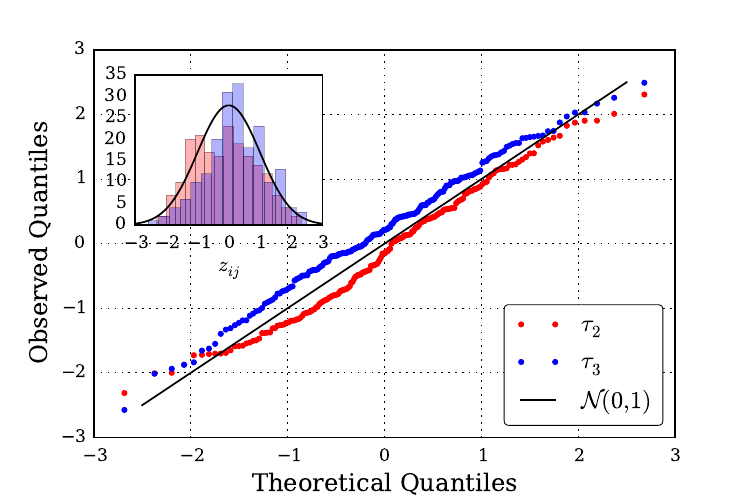}
\caption{\label{fig:standardized}
Quantile-quantile plot of the standardized differences, \cref{eq:zij}, between estimated relaxation timescales, $\tau_2$ and $\tau_3$, on twenty i.i.d. datasets. If the estimated timescales are normally distributed with the calculated asymptotic variances, the quantiles of their standardized differences would match exactly with the theoretical quantiles of the standard normal distribution. }
\end{figure}

How accurate are the approximate asymptotic uncertainty expressions derived in \cref{sect:uncertainty}? To answer this question, we performed a numerical experiment with twenty independent and identically distributed collections of trajectories of Brownian dynamics on a two-dimensional potential. One of those trajectories is shown superimposed on the potential in \cref{fig:muller}, along with the $8 \times 8$ grid used to discretize the process. The Brownian dynamics simulations were performed following the same procedure described in McGibbon, Schwantes and Pande.\cite{McGibbon2014Statistical}

To assess the accuracy of the asymptotic approximations, we compare the empirical distribution of the estimated parameters over the separate data sets with the theoretical distribution which would be expected based on the Gaussian approximation. Consider a scalar model parameter $g$, such as one of the relaxation timescales or equilibrium populations. Fitting a model separately on each of the twenty data sets yields estimates, $\{(\hat{g}_1, \sigma^2_{\hat{g}_1}), \ldots (\hat{g}_{20}, \sigma^2_{\hat{g}_{20}})\}$. If these estimates are accurate, then $\hat{g}$ is normally distributed, $\hat{g} \sim \mathcal{N}(g, \sigma^2_{\hat{g}})$. Our goal is to examine the accuracy of the estimated variances, $\sigma^2_{\hat{g}}$. Note that the true value of $g$ is unknown, but subtracts out when examining standardized differences between the estimates, which, assuming normality, should follow a standard normal distribution,
\begin{align}
\label{eq:zij}
z_{ij} = \frac{\hat{g}_i - \hat{g}_j}{\sqrt{\sigma^2_{\hat{g}_i} + \sigma^2_{\hat{g}_j}}} \overset{?}{\sim} \mathcal{N}(0, 1).
\end{align}

In \cref{fig:standardized}, we compare the empirical and theoretical distributions of $z_{ij}$,  $(i, j) : 1 \leq i \leq 20, i < j \leq 20$, for estimates of the first two relaxation timescales using a quantile-quantile (Q-Q) plot, a powerful method of comparing distributions. The observation that Q-Q plot runs close to the $y=x$ line is encouraging, and shows that the observed deviates are close to normally distributed, and that the approximator's variance estimates are of the appropriate magnitude. This suggests that the asymptotic error expressions can be of practical utility for practitioners.

\subsection{Comparison with discrete-time MSMs}
\label{ss:comp_msm}

\begin{center}
\begin{figure}
\includegraphics[width=3in]{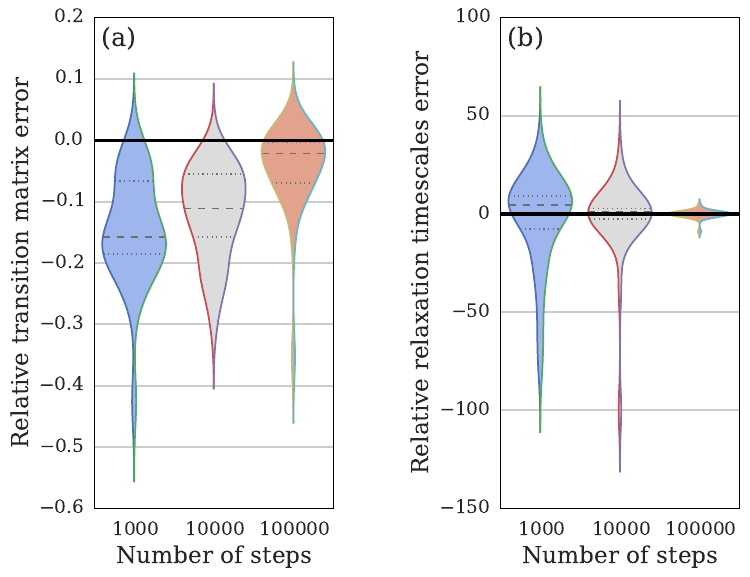}
\caption{\label{fig:conv-vs-msm}Violin plots of the relative error between continuous-time and discrete-time Markov models for kinetics on random graphs. Values below zero indicate lower error for the continuous-time model, whereas values above zero indicate the reverse. The shape displays the data density, computed with a Gaussian kernel density estimator. Panel (a): as measured by the Frobenius-norm error in the estimated transition matrices, $||\mathbf{\hat{T}}-\mathbf{T}||_F$, the continuous-time model achieves lower errors, with a larger advantage for shorter trajectories. Panel (b): as measured by the max-norm error in the estimated relaxation timescales, $\max_i |\hat{\tau}_i - \tau_i|$, the two models are not distinguishable.}
\end{figure}
\end{center}

In a data-limited regime, are continuous-time Markov models more capable than discrete-time MSMs? We extended the analysis in \cref{exp:1} to a larger class of generating processes in order to address this question. We began by sampling random 100-state Markov process rate matrices from scale free random graphs.\cite{Barabasi1999Emergence}. Details of the random rate matrix generation are described in \cref{appendix:randomrates}.

From each random rate matrix, $\mathbf{K}$, we sampled three discrete-time trajectories of different lengths. Each trajectory was used individually to fit both a continuous-time and discrete-time Markov model, and the parameterized models were then compared to the underlying system from which the trajectories were simulated to assess the convergence properties of the approaches.

In \cref{fig:conv-vs-msm}, we consider two notions of error. The first norm measures error in the elements of the estimated transition matrix, $||\hat{\mathbf{T}}-\mathbf{T}||_F$. Unlike the experiment in \cref{fig:convergence}, we used the $\hat{\mathbf{T}}$ as the basis of the measure so that the continuous-time and discrete-time models could be compared on an equal footing. The second error norm we consider is the max-norm error in the estimated relaxation timescales, $\max_i |\hat{\tau}_i - \tau_i|$, which measures a critical spectral property of the models. In both panels of \cref{fig:conv-vs-msm}, the distribution of the difference in error between the continuous-time and discrete-time models is plotted; values below zero indicate that the continuous-time model performed better for a particular class of trajectories, whereas values above zero indicate the reverse. For each condition, we performed $N=30$ replicates.

Our results show that as measured by the transition matrix error, the continuous-time Markov process model is more accurate in the regimes considered.
A binomial sign test rejects the hypothesis that the two estimators give the same error for all three conditions (two-sided $p$ values of $[2 \times 10^{-9}, 2 \times 10^{-9}, 1\times10^{-3}]$ for trajectories of length $[10^3, 10^4, 10^5]$ steps, respectively). The relative advantage of the continuous-time Markov model decreases as the trajectory length increases -- its advantage is in the sparse data regime when no transition counts have been observed between a significant number of pairs of states.

In contrast, as measured by the relaxation timescale estimation error, we observe no significant difference between the continuous-time  and discrete-time estimators. A binomial sign test does not definitively reject the hypothesis that the two estimators give the same error for any of the three conditions (two-sided $p$ values of $[0.02, 0.36, 0.85]$ for trajectories of length $[10^3, 10^4, 10^5]$ steps, respectively). Neither estimator is consistently more accurate in recovery of the dominant spectral properties of the dynamics.

\subsection{Application to Protein Folding and Lag Time Selection}
\label{ss:protein}
\begin{figure}
\includegraphics[width=2.5in]{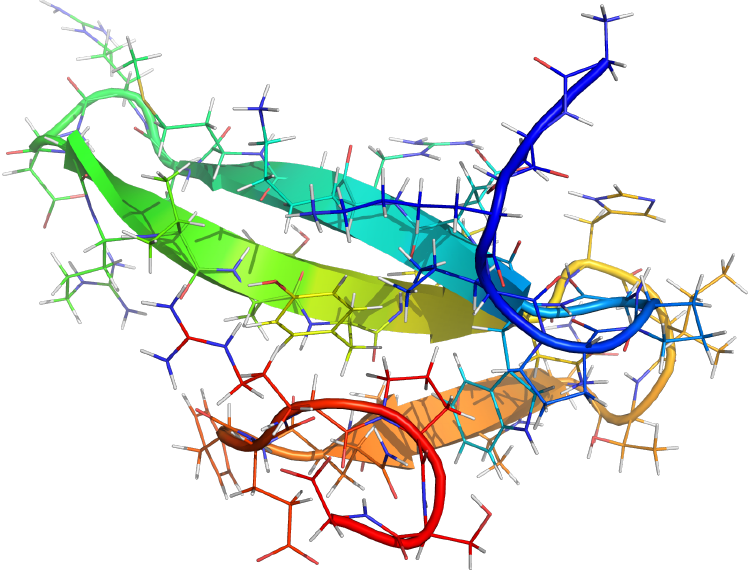}
\caption{\label{fig:ww_rendered} The FiP35 WW protein, in its native state. We analyzed two 100$\mu s$ MD trajectories of its folding performed by D.E. Shaw Research to estimate a Markov process model for its conformational dynamics.\cite{shaw2010atomic}}
\end{figure}
How can these models be applied to the analysis of molecular dynamics (MD) simulations of protein folding? We obtained two independent ultra-long 100$\mu s$ MD simulations of the FiP35 WW protein,\cite{liu2008experimental} a small 35 residue $\beta$-sheet protein (\cref{fig:ww_rendered}), performed by D.E. Shaw Research on the ANTON supercomputer.\cite{shaw2010atomic}

In order to focus on the construction of discrete-state Markov models, we initially projected every snapshot of the MD trajectories, which were available at a 200 $ps$ time interval, into a discrete state space with 100 states in a way consistent with prior work.\cite{McGibbon2014Statistical} Briefly, this involved the extraction of the distance between the closest non-hydrogen atoms in each pair of amino acids in each simulation snapshot,\cite{mcgibbon2014} followed by the application of time-structure independent components analysis (tICA) to extract the four most slowly decorrelating degrees of freedom,\cite{schwantes2013Improvements, perezhernandez2013Identification} which were then clustered into 100 states using the $k$-means algorithm.\cite{lloyd1982least, arthur2007kmeans}

\begin{center}
\begin{figure*}
\centering
\includegraphics[width=\textwidth]{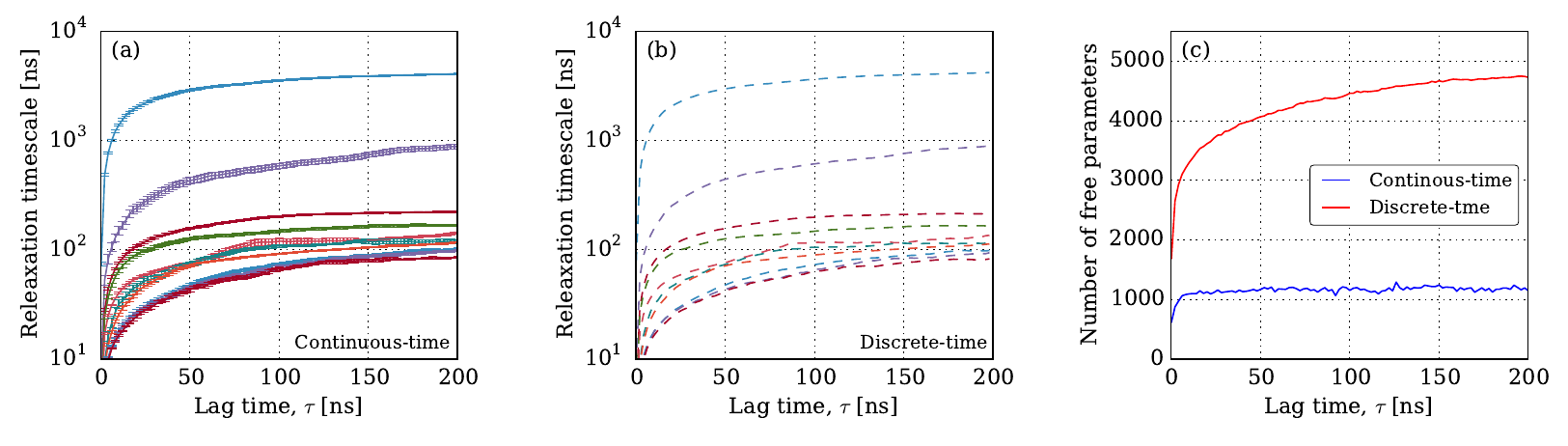}
\caption{\label{fig:ww_timescales} Implied exponential relaxation timescales of parameterized (a) continuous-time Markov process model and (b) discrete-time Markov model as a function of lag time. The relaxation timescales computed by the two algorithms coincide almost exactly ($r^2 = 0.999978$). (c) The number of free (non-zero) parameters estimated by the discrete-time and continuous-time models respectively; the continuous-time Markov model achieves a more parsimonious representation of the data.}
\end{figure*}
\end{center}

\begin{figure}
\includegraphics[width=3in]{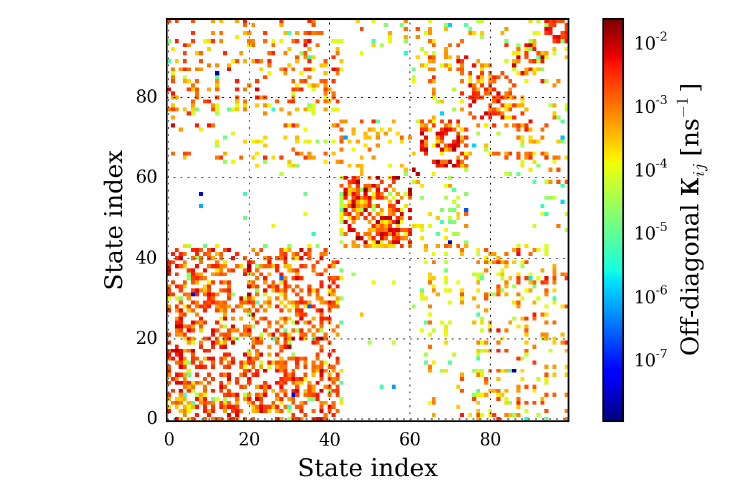}
\caption{\label{fig:ww:ratemat} The maximum likelihood rate matrix, $\hat{\mathbf{K}}$, computed at a lag time, $\tau$, of 100 ns. The state indices were sorted in seven macrostates using the PCCA algorithm.\cite{deuflhard2000identification}}
\end{figure}

\begin{figure}
\includegraphics[width=3in]{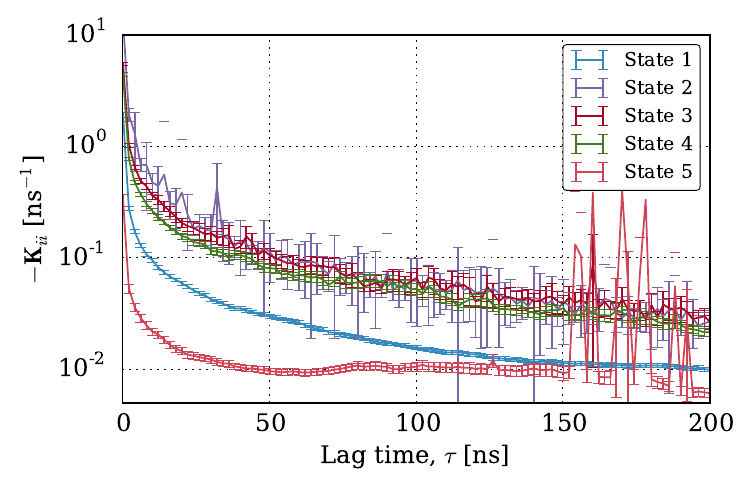}
\caption{\label{fig:ww_lifetimes} Convergence of selected rate matrix elements as a function of lag time. A plausible method for lag time selection would be to choose $\tau$ such that some or all of these entries are determined to have plateaued.}
\end{figure}

Although the equations of motion for a protein's dynamics in an MD simulation are Markovian, the generating process of the data analyzed by our model is not. The pre-processing procedure which projects the original dynamics from a high-dimensional continuous state space (the position and momenta of the constituent atoms) into a lower dimensional continuous space or discrete state space is not information preserving, and destroys the Markov property.\cite{mori1965transport, zwanzig1961memory} For chemical dynamics, qualitative features of the non-Markovianity are well-understood. Consider, for example, a metastable system with two states, $A$ and $B$, the system in state $A$ may stochastically oscillate across the boundary surface many times without committing to state $B$. Whereas for a Markov process, the probability distribution of the waiting time that the system spends in any states before exiting is exponential, chemical dynamics are expected to show a higher propensity for short waiting times, corresponding to so-called recrossing events.\cite{voter1985dynamical} This effect is more pronounced when viewing the process at short lag times -- the bias induced by approximating the process as Markov decreases with lag time.\cite{sarich2010approximation}

For the FiP35 WW domain, we observe that the change in the relaxation timescales of the continuous-time and discrete-time Markov models with respect to lag time are essentially identical, as shown in \cref{fig:ww_timescales}. For both model classes, the estimated relaxation timescales increase and converge with respect to lag time. This is consistent with our results in \cref{fig:conv-vs-msm} (b), which suggest that the estimated timescales are the same for both models, especially as the length of the trajectories grow. While fitting the models in \cref{fig:ww_timescales}, we observed a small number (2-4) of convergence failures at long lag times, which were notable due to a dramatic discontinuinity in the relaxation timescales curve. This problem was solved by reinitializing the optimization at these lag times from the converged solutions at adjacent lag times.

Because of the essentially unchanged nature of the relaxation timescale spectrum, we suggest that when choosing a particular lag time, the same approach be used for discrete-time and continuous-time Markov models. Ideally, this entails the selection of a lag time large enough that the relaxation timescales are independent of lag time.\cite{swope2004describing, Bowman2008using} For the continuous-time Markov model, other techniques may be appropriate as well. For example, in \cref{fig:ww_lifetimes} we show the convergence of selected diagonal entries of the rate matrix as a function of lag time. As described in the context of transition state theory, these rate constants should plateau with increasing $\tau$, which provides another related basis on which select the parameter.\cite{chandler1978plateau, chandler1998barrier}

The most significant difference between the continuous-time and discrete-time estimators in this case is the sparsity of the parameterized models. In \cref{fig:ww_timescales}(c), we compare the number of non-zero independent parameters for both models as a function of $\tau$. Of the $\binom{n+1}{2}=5050$ independent parameters for both the continuous-time and discrete-time models, only $\approx 1200$ are nonzero for the continuous-time model, regardless of lag time. In contrast, the number of nonzero parameters for the discrete-time model model continues to increase with lag time.

We anticipate that the sparsity of $\mathbf{K}$ may aid in the analysis and interpretation of Markov models. In \cref{fig:ww:ratemat}, we show the MLE rate matrix computed at $\tau=100$ $ns$. The state indices were sorted such that states grouped together via Peron cluster cluster analysis (PCCA) were given adjacent indices.\cite{deuflhard2000identification} The evident block structure of the matrix visually indicates that the protein's conformational space consists of a small number of regions with generally high within-region rate constants, but weak between-region coupling. Although a detailed analysis of the biophysics of these conformations is beyond the scope of this work, visual analysis of these structures indicate that the model resolves folded and unfolded, as well as partially folded intermediate states.

In interpreting the $1.96\sigma$ error bars on the relaxation timescales in \cref{fig:ww_timescales}(a), cautionary note is warranted. Our error analysis considers the number of observed transitions between states but does not take into account any notion of uncertainty in the proper definitions of the states themselves, or the error inherent in approximating a non-Markovian process with a Markov process. The observation in \cref{fig:ww_timescales}(a) that the magnitude of the systematic shift in the timescales with respect to lag time is
much larger than the error bars suggests that the Markov approximation (a model misspecification) is a larger source of error, for this dataset, than the statistical uncertainty in model parameters. For these reasons, we caution that these error bars should be interpreted as lower bounds rather than upper bounds.

\subsection{Performance}
In order to assess the performance of our maximum likelihood estimator, we compared it with an algorithm by Holmes and Rubin, which solves the same Markov process parameterization problem using an expectation-maximization approach.\cite{holmes2002expectation} Because the original code was unavailable, we reimplemented the algorithm following the description by Metzner et al., where it is denoted ``Algorithm 4: Enhanced MLE-method for the reversible case''.\cite{Metzner2007353} The algorithm scales as $O(n^5)$, where $n$ is the number of states. Its rate limiting step involves an $O(n^5)$ FLOP contraction of five $n \times n$ matrices into a rank-4 tensor of dimension $n$ on each axis. \footnote{Both our algorithm and Holmes-Rubin estimator were implemented the Cython language and compiled to C++. Our implementation of the Holmes-Rubin estimator is available at \url{https://github.com/rmcgibbo/holmes_rubin}} For benchmarking, we constructed a variant of the the FiP35 WW protein dataset from \cref{ss:protein}, in which we varied the number of states between 10 and 100 during clustering. All models were fit on an Intel Xeon E5-2650 using a single CPU core.

\begin{figure}
\includegraphics[width=3in]{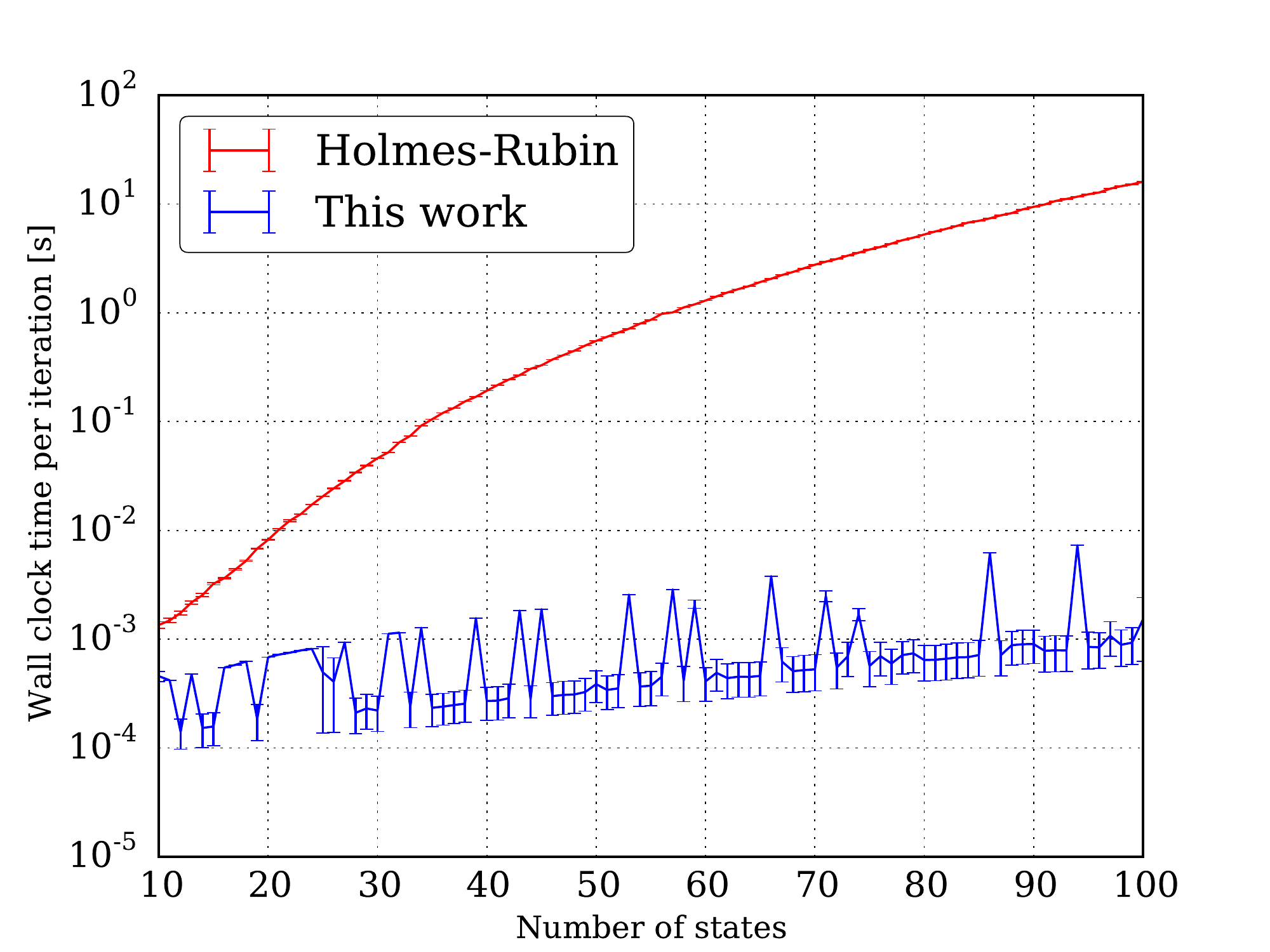}
\caption{\label{fig:performance} Performance of our Markov process estimator, as compared to the Holmes-Rubin EM estimator.\cite{holmes2002expectation} Each iteration of our $O(n^3)$ estimator takes on the order of 1 ms, while the $O(n^5)$ Holmes-Rubin estimator takes over 10 seconds per iteration for a 100 state model. Using default convergence criteria, our estimator often achieves a solution long before the EM estimator finishes a single iteration.}
\end{figure}

As shown in \cref{fig:performance}, and expected on the basis of the $O(n^5)$ vs. $O(n^3)$ scaling, the performance difference between the algorithms is substantial. For $n=100$, our algorithm is roughly four orders of magnitude faster per iteration; our algorithm takes on the order of 1 ms per iteration, while the Holmes-Rubin estimator's iteration takes over 10 seconds. Using the L-BFGS-B optimizer's default convergence criteria, roughly three quarters (68/91) of the runs of our algorithm converge in fewer than 100 iterations; a solution is often achieved long before the EM estimator has performed a single iteration.

\section{Conclusions}

In this work, we have introduced a maximum likelihood estimator for continuous-time Markov processes on discrete state spaces. This model can be used to estimate transition rates between various substates in a dynamical system based on observations of the system at a discrete time interval. Various constraints on the solution, such as detailed balance, can be easily incorporated into the model, and asymptotic error analysis can give confidence intervals in model parameters and derived quantities.

With the efficient parameterization problem solved, these continuous-time Markov models offer several advantages over existing MSM methodologies. As compared to discrete-time MSMs, these models are more interpretable for chemists and biologists because they do not arbitrarily discretize time. Although a lag time is used internally during parameterization, the final estimated quantities are familiar rate constants from chemical kinetics, as opposed to the somewhat unintuitive transition probabilities in a discrete-time MSM. Furthermore, these models are more parsimonious, and unlike the discrete-time MSM are able to detect that many pairs of states are not immediately kinetically adjacent to one another. This makes it possible to more clearly recover the underlying graph structure of the kinetics. For applications such as the determination of transition pathways in protein dynamics, we anticipate that this property will be valuable.

Many extensions of this model are possible in future work. The simple nature of the constraints on $\theta$ make Bayesian approaches, especially Hamiltonian Monte Carlo, particularly attractive.\cite{neal2011mcmc} In particular, because of the separation of $\theta^{(\pi)}$ and $\theta^{(S)}$ in the parameterization, strong informative priors on $\pi$ may be added to extend the work of Trendelkamp-Schroer and No\'{e}.\cite{Trendelkamp2013efficient} The appropriate sparsity inducing priors on $\theta^{(S)}$ may be a topic of future work.

An implementation of this estimator is available in the MSMBuilder software package at \url{http://msmbuilder.org/} under the GNU Lesser General Public License.

\section*{Acknowledgements}
This work was supported by the National Science Foundation and National Institutes of Health under Grant Nos. NIH R01-GM62868, NIH S10 SIG 1S10RR02664701, and NSF MCB-0954714. We thank Mohammad M. Sultan, Matthew P. Harrigan, John D. Chodera, Kyle A. Beauchamp, Ariana Peck, and the reviewers for helpful feedback during the preparation of this work.

\appendix

\section{Random rate matrices}
\label{appendix:randomrates}
Scale-free random graphs with 100 states were generated using the Barab\'{a}si–Albert preferential attachment model with $m=3$.\cite{Barabasi1999Emergence} From the graph's adjacency matrix, we generated a symmetric rate matrix $S$ by sampling a log-normally distributed
random variable ($\mu=-3$, $\sigma=2$) for each connected edge. The stationary distribution, $\pi$, was sampled from $\mathrm{Dirichlet}(\alpha=1)$. The matrix $S$ was then scaled by $50 \cdot (\sum_{ij}S_{ij})^{-1}$, which tuned the relaxation timescales in the range between $10^2$ and $10^3$ time steps, and used with $\pi$  in \cref{eq:kij} to construct $\mathbf{K}$.

\section*{References}
\bibliography{bibliography}
\end{document}